\documentclass[useAMS,usenatbib,usegraphicx]{mn2e}

\newcommand{\be}{\begin{equation}}
\newcommand{\ee}{\end{equation}}
\newcommand{\bea}{\begin{eqnarray}}
\newcommand{\eea}{\end{eqnarray}}

  \newcommand{\bc}{\begin{center}}
  \newcommand{\ec}{\end{center}}
  \newcommand{\Msun}{~{\rm M_\odot}}
  \newcommand{\hMsun}{~h^{-1}\>{\rm M_\odot}}
  
  \newcommand{\Kpc}{~h^{-1}~{\rm kpc}}

  \newcommand{\vel}{\,{\rm km\,s^{-1}}}
  \newcommand{\sbu}{\mathrm{mag ~ arcsec^2}}
  \newcommand{\pix}{\mathrm{pixel}}

\title[Characterizing diffused stellar light]{Characterizing diffused stellar light in simulated galaxy clusters}

\author[Weiguang Cui, et al.]
{\parbox{\textwidth}{Weiguang Cui,$^{1,2}$\thanks{E-mail: \texttt{wgcui@oats.inaf.it}}
G. Murante,$^{2}$
P. Monaco,$^{1,2}$
S. Borgani,$^{1,2,3}$
G.L. Granato,$^{2}$
M. Killedar,$^{1,2}$
G. De Lucia,$^{2}$
V. Presotto$^{2}$ and
K. Dolag$^{4,5}$}\vspace{0.4cm}\\
\parbox{\textwidth}{$^{1}$Astronomy Unit, Department of Physics,
  University of Trieste, via Tiepolo 11, I-34131 Trieste, Italy\\
$^{2}$INAF, Astronomical Observatory of Trieste, via Tiepolo 11,
I-34131 Trieste, Italy\\
$^{3}$INFN --Sezione di Trieste, via Valerio 2,
I-34100 Trieste, Italy\\
$^{4}$ University Observatory Munich, Scheinerstr. 1, D-81679
Munich, Germany\\
$^{5}$ Max-Planck-Institut fr Astrophysik, Karl-Schwarzschild
Strasse 1, Garching bei Mnchen, Germany
}}

\begin{document}


\pagerange{\pageref{firstpage}--\pageref{lastpage}} \pubyear{2013}

\maketitle

\label{firstpage}

\begin{abstract}

  In this paper, we carry out a detailed analysis of the performance
  of two different methods to identify the diffuse stellar light in
  cosmological hydrodynamical simulations of galaxy clusters. One
  method is based on a dynamical analysis of the stellar component,
  which separates the brightest central galaxy (BCG) from the stellar
  component not gravitationally bound to any galaxy, what we call
  'diffuse stellar component' (DSC). The second method is closer to
  techniques commonly employed in observational studies. We generate
  mock images from simulations, and assume a standard surface
  brightness limit (SBL) to disentangle the BCG from the intra-cluster
  light (ICL). Both the dynamical method and the method based on the
  SBL criterion are applied to the same set of hydrodynamical 
  simulations for a large sample of about 80 galaxy clusters. We analyse
  two sets of radiative simulations: a first set includes the effect
  of cooling, star formation, chemical enrichment and galactic
  outflows triggered by supernova feedback (CSF set); a second one
  also includes the effect of thermal feedback from active galactic
  nuclei triggered by gas accretion on to supermassive black holes (AGN
  set).

  We find significant differences between the ICL and DSC fractions
  computed with the two corresponding methods, which amounts to about
  a factor of 2 for the AGN simulations, and a factor of 4 for
  the CSF set. We also find that the inclusion of AGN feedback boosts the
  DSC and ICL fractions by a factor of 1.5-2, respectively, while
  leaving the BCG+ICL and BCG+DSC mass fraction almost unchanged. The sum
  of the BCG and DSC mass stellar mass fraction is found to decrease 
  from $\sim 80$ per cent in galaxy groups to $\sim 60$ per cent in
  rich clusters, thus in excess of that found from observational
  analysis.   

  We identify the average SBLs that yield the
  ICL fraction from the SBL method close to the DSC fraction from the
  dynamical method. These SBLs turn out to be
  brighter in the CSF than in the AGN simulations. This is consistent with the
  finding that AGN feedback makes BCGs to be less massive and with
  shallower density profiles than in the CSF simulations.  The BCG
  stellar components, as identified by both methods, are slightly older
  and more metal-rich than the stars in the diffuse component. Relaxed
  clusters have somewhat higher stellar mass fractions in the diffuse
  component. The metallicity and age of both the BCG and diffuse
  components in relaxed clusters are also richer in metals and older.
\end{abstract}



\section{Introduction}
\label{i}

The concept of intra-cluster light (ICL)
was first introduced by \cite{Zwicky1951}, who pointed out the
existence of stars between galaxies within the Coma Cluster. However,
the relevance of this baryonic component remained unclear until the
advent of CCD photometry, that allowed accurate measurements in the
Coma cluster to be carried out \citep{Bernstein1995}. Since then,
several papers investigated the ICL in the Coma Cluster
\citep{Gregg1998, Trentham1998}, in the Fornax Cluster
\citep{Theuns1997}, in the Virgo Cluster \citep{Durrell2002,
  Feldmeier2004, Aguerri2005, Mihos2005, Rudick2010}, in other nearby
clusters \citep{Lin2004, Gonzalez2005, Krick2007}, and also at higher
redshift \citep{Zibetti2005, Toledo2011, Guennou2012, Burke2012}.The
ICL is now established to have low surface brightness with smooth
distribution around the central galaxy, extending to large
radii. However, the identification of this ICL in observations remains
difficult and uncertain. Indeed, the typical surface brightness of
the ICL is less than $1$ per cent of the dark sky; also it can be
contaminated by foreground and background galaxies; moreover, it is
difficult (and somewhat arbitrary) to establish where the ICL begins
and the associated brightest central galaxy (BCG) ends.

The kinematic information on the ICL can be obtained by discrete objects
like planetary nebulae (PNe), which are excellent tracers for
measuring the line-of-sight kinematics of intracluster stars
\citep{Arnaboldi2010}. With a suitable number of PNe, the mean radial
velocities and velocity dispersions of the ICL can be determined. Such
investigations have been carried out in a number of clusters
\citep[e.g.][]{Ciardullo2005, Gerhard2007, Doherty2009, Arnaboldi2012}

On the theoretical side, in the last few years the analysis of
simulated clusters provided an important contribution towards
elucidating the nature and origin of the diffuse light component
\citep[e.g.][]{Murante2004, Murante2007, Willman2004, Tutukov2007,
  Sommer-Larsen2005, Puchwein2010, Dolag2010, Rudick2006, Rudick2009,
  Rudick2011, Barai2009, Martel2012}, which is now believed to contain
valuable information on the dynamical history of the cluster and of
the member galaxies. Diffuse light is also relevant for theoretical
investigations of the evolution of the galaxy stellar mass function
\citep{Monaco2006, Yang2009} and of the stellar mass of satellite galaxies
\citep{Liu2010, Kang2008, Conroy2007}. Generation of diffuse light
associated to mergers, along the hierarchical assembling of cluster
galaxies, has been introduced in recent semi-analytic galaxy formation
models \citep{LoFaro2009, Guo2011, Somerville2008, Contini2013}. In
theoretical studies, mostly based on simulations, the existence of an
important component of stars that are not gravitationally bound to any
galaxy, has been pointed out. While this dynamically distinct
component is clearly related to, and often identified with, the
observed ICL, it is not conceptually identical and cannot be directly
compared with it. In the following, it will be dubbed as Diffuse
Stellar Component (DSC). Clearly, it is of paramount importance to
understand the relationship between the two. Careful analysis of
simulations of clusters that model the formation of stars in galaxies
is the most natural way to make progresses on this issue. This is the
main aim of this paper.

A basic question about the ICL/DSC is what fraction of cluster stars are
in these components. To date, consensus has not been reached yet on
this issue either from observations or from simulations. This is
attributable to some extent to the different adopted criteria to
define and/or identify the ICL and the DSC. Observational estimates cover the
range from $2$ to $50$ per cent. \cite{Lin2004} find a
positive correlation of the ICL fraction with halo mass, reaching
$\sim 50$ per cent for clusters of $10^{15} \Msun$. \cite{Zibetti2005}
find that the ICL fraction is independent of cluster richness, with an
average value of $\sim 10$ per cent. \cite{Krick2007} note a large
scatter in the ICL fraction from 6 to 20 per cent, but without any
trend with mass. However, \cite{Sand2011} claim a declining stellar
mass fraction as a function of halo mass, as reported also by
\cite{Gonzalez2007}. Similarly, theoretical studies have sometimes
found no significant dependence of the ICL/DSC fraction on cluster mass
\citep{Dolag2010, Puchwein2010, Rudick2011, Contini2013}, while in
other cases the ICL/DSC fraction has been found to increase with
cluster mass \citep{Purcell2007, Murante2007, Henriques2008,
  Martel2012, Watson2012}. Contrasting claims have been made also on
the dependence of the result on the identification method
\citep{Puchwein2010, Rudick2011}.

In this paper we use and improve two methods to identify the DSC and
the ICL, respectively, in simulations (Section \ref{methods}). The dynamical method
(dubbed {\small SUBFIND} from the name of the substructure identification
algorithm; \citealt{Springel2001,Dolag2009}) is based on exploiting
dynamical information available for star particles in simulations, so
as to identify those particles belonging to the DSC. The SBL method
(dubbed {\small MAP} algorithm) mimics typical observational procedures and is based on
the generation of surface brightness maps, on which a threshold
surface brightness criterion is imposed to identify the ICL. We apply
these methods to two sets of radiative simulations of galaxy clusters
and groups, based on including only the effect of galactic outflows
triggered by supernova (SN) feedback and also adding the effect of
thermal feedback from active galactic nuclei (AGN), respectively.  The
plan of the paper is as follows. We introduce the simulations in
Section \ref{simulation}, while we describe in Section \ref{methods}
the methods for the DSC/ICL identification. In Section \ref{results} we
present our results on the relationship between the ICL and the DSC identified
with the two methods, and on the effect of the AGN feedback on these
components. Our main results are summarized and discussed in Section
\ref{discussion}.

\section{The set of simulated galaxy clusters}
\label{simulation} 

We summarize here the basic features of the set of
simulated clusters analysed in this work. For more detailed
information on the generation of initial conditions and basic
properties of the simulated clusters, we
refer the reader to \cite{Fabjan2011, Bonafede2011}.

The simulations have been carried out with the TreePM-SPH {\small
  GADGET-3} code, a more efficient version of {\small GADGET-2}
\citep{Springel2005}, which includes a scheme of domain decomposition
allowing an improved workload balance, especially in simulations, like
those considered here, where the computational cost is dominated by
few prominent non-linear structures.

The clusters are extracted from high resolution re-simulations of 29
Lagrangian regions, taken from a large volume, low-resolution $N$-body
cosmological simulation \citep[see more details
in][]{Bonafede2011}. The parent simulation follows $1024^3$ DM
particles within a periodic box of comoving size $1 h^{-1} {\rm
  Gpc}$, assuming a flat $\Lambda$CDM model: matter density parameter
$\Omega_{\rm m} = 0.24$; baryon density parameter $\Omega_{\rm b} = 0.04$; Hubble
constant $h = 0.72$; normalization of the power spectrum $\sigma_8 =
0.8$; primordial power spectral index $n_s = 0.96$. Each Lagrangian
region has been re-simulated at higher resolution employing the {\it
  Zoomed Initial Conditions} (ZIC) technique \citep{Tormen1997}. Each
high-resolution Lagrangian region is taken to be large enough so that
no contaminant low-resolution particle is found at $z=0$ within
five virial radii of the central cluster. Given the resulting large
size of the high-resolution regions, each of them contains more than
one galaxy group or cluster not contaminated by low-resolution
particles within the virial region. Gas particles are added only in
the high-resolution regions, where DM and gas particles have masses
of $8.47\times 10^8\hMsun$ and $1.53\times 10^8\hMsun$ respectively. The
Plummer--equivalent gravitational softening is fixed to $5\,h^{-1}$kpc
in comoving units below $z=2$, while being fixed in physical units at
higher redshift. The B-spline smoothing length for the SPH
computations is allowed to reach a minimum value of half of the
gravitational softening. 

All the simulations analysed in this paper include radiative gas
cooling, star formation and chemical enrichment. Radiative cooling
rates are computed following the same procedure by \cite{Wiersma2009},
by accounting for the contribution to cooling from 11 elements (H,
He, C, N, O, Ne, Mg, Si, S, Ca, Fe), provided by the CLOUDY
photoionization code \citep{Ferland1998} for an optically thin gas in
photoionization equilibrium, also accounting for the presence of the
cosmic microwave background and of UV/X-ray background radiation from
quasars and galaxies \citep{HaardtMadau2001}.  Star formation is
described through the sub-resolution model originally introduced by
\cite{Springel2003}. In this model each gas particle exceeding a
threshold density of $n_{\rm H}=0.1 {\rm cm}^{-3}$ is assumed to be
multiphase and star forming, with a hot ionized phase coexisting in
pressure equilibrium with a cold phase which describes the
interstellar medium and provides the reservoir for star formation. In
addition to the density criterion, we also introduce a threshold
temperature of $2.5\times 10^5$K, below which a gas particle can be
treated as multiphase. Chemical enrichment is included by following
the production of heavy elements from SN-II, SN-Ia and low- and
intermediate- mass stars, as described by \cite{Tornatore2007}. Stars
of different mass, distributed according to a Chabrier IMF
\citep{Chabrier2003}, release metals over the time-scale determined
by their mass-dependent lifetimes, taken from
\cite{PadovaniMatteucci1993}.

In this paper, we will focus on two sets of such cluster
simulations. Besides the effect of radiative cooling and star
formation, a first set contains cooling, star formation and the effect
of kinetic feedback through galactic outflows triggered by SN-II
explosions (CSF set hereafter). In the model of kinetic feedback,
originally introduced by \cite{Springel2003}, a multi-phase
star-forming gas particle is assigned a probability to be uploaded in
galactic outflows, which is proportional to its star formation
rate. We assume $v_w = 500\vel$ for the outflow velocity and
a mass-upload rate that is two times the value of the star
formation rate of a given particle.

Besides kinetic feedback, a second set of simulations also include the
effect of a thermal AGN feedback (AGN set hereafter). The model of
this AGN feedback is largely inspired to the original
implementation by \cite{Springel2005b}, but with some significant
changes, as described in detail by \cite{Cin2013}.  Super massive
black holes are initially seeded as sink particles, with an
initial mass of $5\times 10^6\hMsun$, within haloes identified by a
friend-of-friend algorithm which have a minimum mass of $2.5\times
10^{11}\hMsun$ and do not already include a black hole particle. Super
massive black holes grow in mass by merging with other black holes and
via gas accretion that proceeds at an Eddington--limited Bondi
rate. Differently from the original implementation by
\cite{Springel2005b}, the black hole mass increase by gas 
accretion does not proceed through a stochastic swallowing of gas
particles. Instead, the dynamical mass of a black hole particle continuously
increases according to the accretion rate, without removing the
surrounding gas. This prescription causes a minor violation of mass
conservation. However, it improves the numerical stability of the
accretion process and prevents accretion from affecting the gas
distribution on scales that are physically implausible because of the
limited numerical resolution (\citealt{Cin2013}; see also
\citealt{Wurster2013}). A radiative efficiency parameter,
$\epsilon_r$, describes the fraction of the accreted rest-mass energy
that is radiated from the black hole, while a feedback efficiency parameter,
$\epsilon_f$, corresponds to the fraction of this radiated energy that
is thermally coupled to the surrounding gas. We further assume that
feedback enters in a quiescent radio mode whenever the accretion rate
drops below one-hundredth of the Eddington limit. In this regime,
feedback efficiency is increased by a factor of 4 \citep[see
also][]{Sijacki2007,Fabjan2010}. As shown in \cite{Cin2013}, our
implementation of the AGN feedback allows us to reproduce the observed
relation between black hole masses and stellar masses of host galaxies for the
adopted values of $\epsilon_r=0.2$ and $\epsilon_f=0.2$.

\section{The Diffuse Light Identification Methods}
\label{methods}

In observations, two methods are usually employed to identify the
ICL. The first one is to attribute to the ICL
all the light coming from outside some optical boundaries of galaxies,
usually defined by a fixed surface brightness limit (SBL)
\citep[e.g.][]{Feldmeier2004, Mihos2005, Zibetti2005}. The second
method is to separate the BCG from ICL by modeling the
surface brightness with two profiles (usually de Vaucouleurs or
Sérsic), after masking satellite galaxies and foreground contamination
\citep[e.g.][]{Gonzalez2005, Seigar2007}. In this paper, we apply the
first method, after generating mock cluster images from the
simulations, as described in Section \ref{map} \citep[see
also][]{Rudick2011}. 
In simulations, it is possible to exploit the full dynamical
information available to investigate the
DSC \citep[e.g.][]{Murante2007, Rudick2011, Puchwein2010,
  Dolag2010}, which is defined by the stars gravitationally bound to
the gravitational potential of a cluster but not to that of any galaxy
\citep[e.g.]{Dolag2010}. There are several algorithms that analyse the
kinematics of star particles and separate the DSC from the BCG \citep[][and
references therein]{Murante2004, Puchwein2010, Dolag2010}. In this
paper, we use a slightly modified version of {\small SUBFIND}
\citep{Dolag2010}, which we describe in detail in Section
\ref{subfind}. We refer the first method as SBL
method which applies an algorithm named {\small MAP}. While the second
method is called dynamical method which uses the {\small SUBFIND} algorithm.

Although the ICL and the DSC are clearly related to each other, in general
they do not include the same population of stars. Comparing their
properties is quite interesting for a number of reasons, such as
inferring the dynamical origin of the ICL, whose identification is
based on observational criteria not related to the kinematics of
stars.

\begin{figure}
\includegraphics[width=0.5\textwidth]{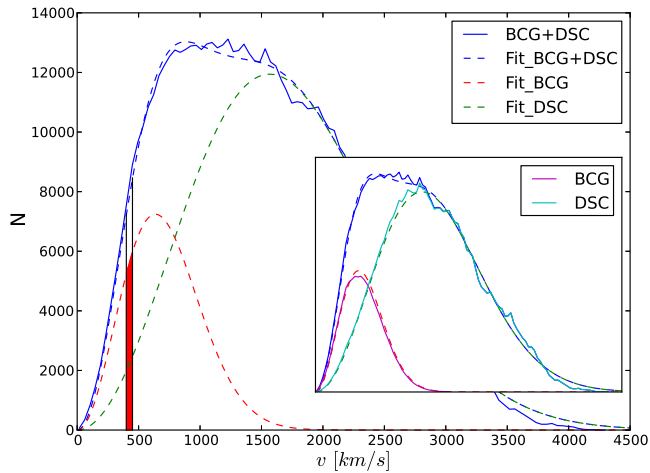}
\caption{Velocity histograms of the BCG and DSC stellar particles
  (blue solid line) and the corresponding double Maxwellian fit (blue
  dashed line).  The red and green dashed lines are the two single
  Maxwellian fitting profiles. The two vertical lines indicate a
  velocity bin, whose red area represents the stars assigned to the
  BCG with the criterion based on potential energy described in
  Section \ref{subfind}. The solid magenta and cyan lines in the small
  inset panel show the velocity histograms of the BCG and DSC stars
  after the separation process, along with the corresponding
  individual single-Maxwellian fitting curves.}
\label{fig:cartoon}
\end{figure}

\begin{figure}
\includegraphics[width=0.5\textwidth]{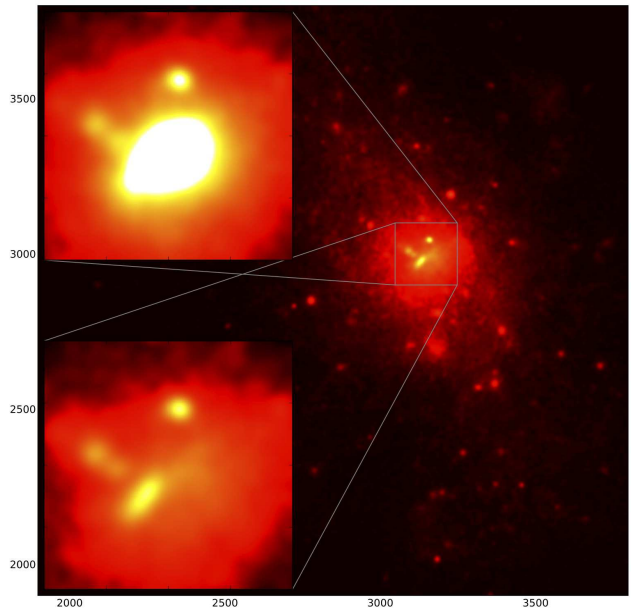}
\caption{Projected density map of the DSC in one simulated cluster at
  $z=0$. The lower and upper zoomed-in panels show the centre of the
  cluster without and with the BCG, respectively. Units of the axes
  are in $\Kpc$.}
\label{fig:image}
\end{figure}

\subsection{The dynamical method}
\label{subfind}

The {\small SUBFIND} algorithm \citep{Springel2001, Dolag2009}
algorithm was originally introduced to identify substructures within a
group of particles identified by applying the friends-of-friends (FoF)
algorithm. Afterwards, the {\small SUBFIND} algorithm was suitably
modified to separate the DSC from the BCG by \cite{Dolag2010}.
{\small SUBFIND} identifies 
the main sub-halo as the most massive gravitationally bound
substructure, which includes both the BCG and DSC. The velocity
dispersion distribution of star particles inside this main sub-halo is
best described by a double Maxwellian \citep[see more details
in][]{Dolag2010, Puchwein2010}. It is then natural to assign the lower
velocity dispersion component to the BCG, while the higher velocity
dispersion component is assumed to feel the general potential of the
cluster and, as such, is identified with the DSC.

While this statistical procedure provides an estimate of the stellar
mass fraction in the BCG and DSC components, it does not allow one to
assign individual star particles to either one of the two
components. On the other hand, this is required for most of the
analysis presented in this paper, and is implemented by means of an
iterative analysis \citep{Dolag2010,Puchwein2010}. In brief, a radius
is iteratively searched, such that the velocity distribution of
particles which are gravitationally bound to the matter contained
within that radius; matches the lower dispersion Maxwellian
component. Those star particles are defined to belong to the BCG.

Here, we also employ an analysis of binding energies, but with a novel
algorithm that is both simpler and does not require iterations. 
Furthermore, this new algorithm can avoid mis-matching between the
fitted velocity dispersion profiles and these from the simulated BCG
particles and DSC particles. In
each velocity bin, we assign to the BCG all the star particles whose
potential is lower than a limiting value, defined so that the mass
fraction below it matches the ratio between the area below the lower
Maxwellian within a given velocity bin (red region in Fig.
\ref{fig:cartoon}) and the area below the total velocity distribution
encompassed by the same bin. The remaining particles are assigned to
the DSC. In this procedure, the DSC and BCG are allowed to
be spatially overlapping,
as they do in the double Maxwellian fit. In this way the two
components provide an exact fit to the two Maxwellian 
profiles, without the need of an iteration process which could 
suffer from convergence problems. Moreover, it is physically
reasonable to assign lower potential particles to the BCGs.

In order to give a visual impression of how this method separates the
two components, we show in Fig. \ref{fig:image} the projected image
of the DSC and BCG of the same simulated cluster. Colour coding is
based on the projected surface density of star particles. Quite
clearly, removing the BCG components provides a rather smooth
distribution of the remaining diffuse component.

\subsection{The SBL method}
\label{map}

To generate surface brightness maps of the clusters, we follow the
same procedure as in \cite{Cui2011}. Each star particle of the FoF
group is treated as a Simple Stellar Population (SSP) with age,
metallicity and mass given by the corresponding particle's properties
in the simulation, and adopting the same initial mass function (IMF)
\citep[a Chabrier IMF;][]{Chabrier2003}. The spectral energy
distribution (SED) of each particle is computed by interpolating the
SSP templates of \cite{Bruzual2003}. A standard Johnson $V$-band filter
is applied to this SED to calculate its $V$-band luminosity. Then, we
smooth to a 3D mesh grid both the luminosity and the mass of each star
particle with the same spline kernel used for the SPH
calculations. The mesh size is fixed to $5 \Kpc$ (corresponding to the
gravitational softening length), and we use 64 SPH neighbours. The
exact value of these parameters does not significantly affect our final
results \citep[see also][]{Rudick2006}. Finally, projecting this 3D
mesh in one direction yields the 2D photometric image. Since only the
star particles within the FoF group are used, we take out all the
contamination from the background or foreground galaxies, that might
significantly affect the observational results. We neglect dust
reprocessing in this procedure, which should have less effect in the
dust-free clusters. To measure the ICL in the mock
images, we use the SBL usually adopted in observations $\mu_V
> 26.5~ \sbu$ \citep[e.g.][]{Mihos2005, Feldmeier2002,
  Rudick2011}. The choice of a SBL of $\mu_V > 26.5$ for the SBL
method is motivated by the fact that this surface brightness is used
to define the Holmberg radius \citep{Holmberg1958}, a commonly adopted
method to define the isophotal size of galaxies. This SBL is also used
to mark the transition where the ICL begins to take on
a distinct morphology from the galactic light in both simulations
\citep{Rudick2006, Rudick2009, Rudick2011} and observations
\citep{Feldmeier2004, Mihos2005, Rudick2010b}. However, applying this
simple SBL cut to clusters of various masses and dynamical states, may
not be able to capture individual BCGs.

To understand the role of the SBL in the BCG/ICL separation, we also
apply two other cuts, one magnitude brighter and fainter
than the above reference value.  To define the mass of the BCG, we
draw circular annuli centred on the pixel containing the most bound
particle with width $dR = 1.5~ \pix$. We then define the outer
boundary of the BCG as the innermost annulus in which the fractional
area with surface brightness fainter than $26.5~ \sbu$ is greater than
0.5, and we attribute to BCGs the integrated flux of all the
pixels (with $\mu_V < 26.5~\sbu$) inside this
radius. \cite{Rudick2011} also applied a similar procedure to generate
surface brightness maps of simulated clusters. However, their
procedure is based on a dark matter only simulation.

\section{Results}
\label{results}

\begin{figure*}
\includegraphics[width=1.0\textwidth]{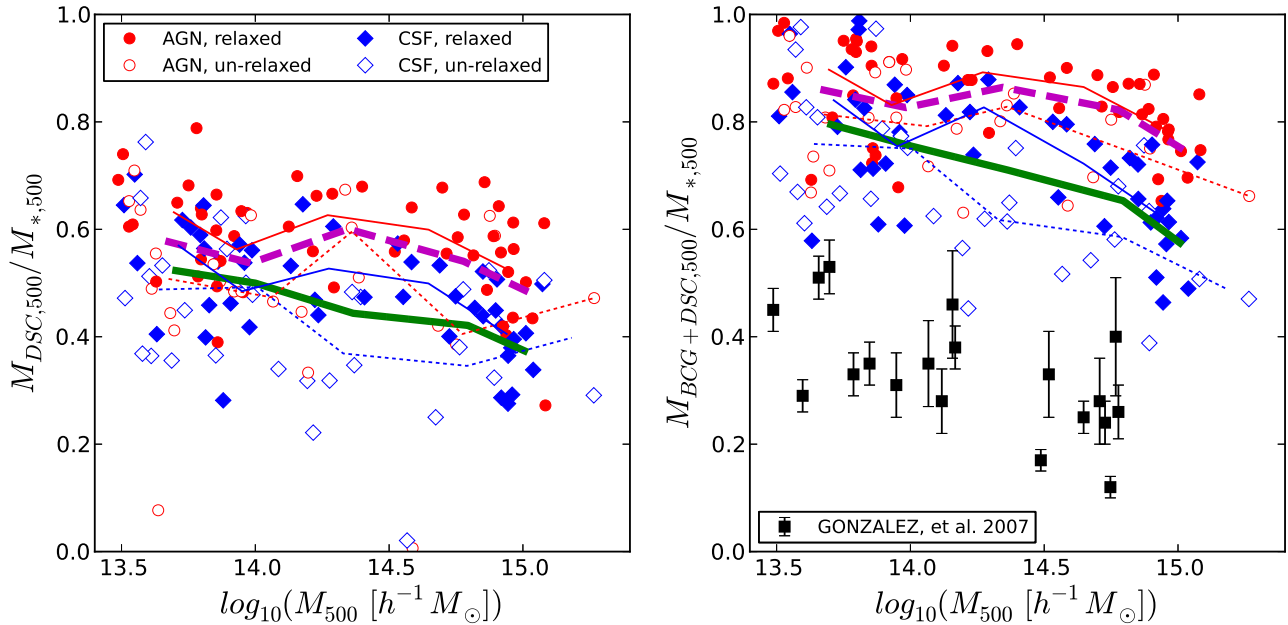}
\caption{Left panel: The DSC fraction within $r_{500}$, computed with the
  dynamical method, as a function of halo mass $M_{500}$. Right panel:
  as in the left panel but for the BCG+DSC fraction. Blue squares and
  red circles represent the CSF and AGN simulations, respectively. The
  solid green and the dashed magenta thick lines are the corresponding
  averages. We have used filled and open symbols to denote relaxed and
  un-relaxed clusters. The thin solid and dotted lines are the
  averages for filled and open symbols, respectively.} \label{fig:ICLf_L}
\end{figure*}

\begin{figure*}
\includegraphics[width=1.0\textwidth]{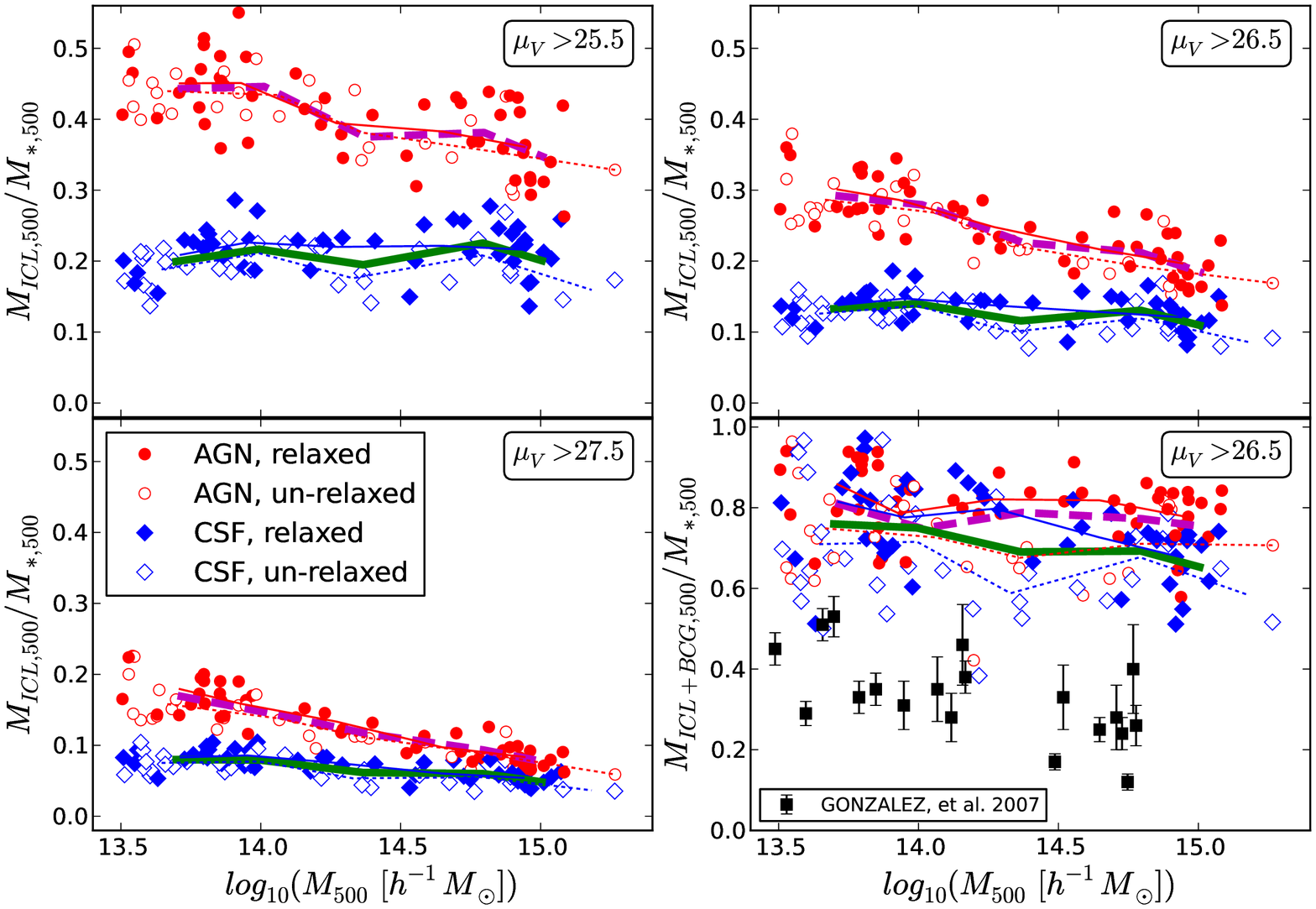}
\caption{The ICL mass fraction within $r_{500}$, from the SBL method, as a
  function of halo mass $M_{500}$ . The upper panels and the lower
  left panel show the ICL fractions computed assuming different values
  of the SBL, as indicated in each panel (in unit
  of $\sbu$).  The symbols and lines have the same meanings as Fig.
  \ref{fig:ICLf_L} for the relaxed and un-relaxed clusters. The lower
  right panel shows the BCG+ICL fraction with a SBL of $\mu_V > 26.5
  \sbu$.}
\label{fig:ICLf_M}
\end{figure*}

\subsection{The ICL-DSC fraction}
\label{IDfrac}

As already discussed in Section \ref{i}, the fraction of the ICL and
the DSC
have been investigated in previous analyses, both in simulations and
in observations, with results sometimes discordant. This can be partly
attributed to the different methods and definitions adopted to
identify these components.  In the following, all the fractions of
stars in one of these components are computed within $r_{500}$, which
is defined as the radius within which the mean matter density is $500$
times the critical cosmic density.

In the left panel of Fig. \ref{fig:ICLf_L}, we show the fraction of
the total stellar mass in the DSC, as a function of halo mass
$M_{500}$. We have used filled and open symbols to distinguish between
relaxed and un-relaxed clusters, defined following the method of
\cite{Killedar2012}. This method consists in computing the offset
between the minimum potential position and centre of mass, calculated
within a range of radii $\zeta_i r_{vir}$, with $\zeta_i$ going from 
$0.05$ to $2$ in 30 logarithmic steps. A cluster is defined as relaxed
if the offset is less than 10 per cent of $\zeta_i r_{vir}$ for all
radii. 

For the CSF set, the DSC mass fraction turns out to vary from $\sim
50$ to 40 per cent when going from lower to higher halo mass, with
fairly large scatters around these values. This fraction is $\sim 55$
per cent for the AGN simulations, almost independent of the halo mass.
The DSC fraction for the CSF set is consistent with that found
by\cite{Puchwein2010}, and slightly higher than that reported by
\cite{Dolag2010}. Furthermore, the DSC fraction is about $10$ per cent
higher for the AGN set, at variance with the claim of
\cite{Puchwein2010} that this fraction does not significantly change
with and without AGN feedback. Relaxed clusters tend to have higher
DSC fractions than un-relaxed clusters. Relaxed clusters should be
dynamical evolved systems, compared to the clusters under
merging. Thus, a higher DSC fraction in relaxed clusters can be
understood if the diffuse star particles mainly come from the
satellite galaxies undergoing merging and tidal stripping. We checked
that the DSC fractions within $r_{200}$ have a very similar behaviour.

In the right panel of Fig. \ref{fig:ICLf_L} we show the BCG+DSC mass
fractions within $r_{500}$.  With the inclusion of the BCG mass, these
fractions boost to $\sim 60 - 80$ per cent for the CSF case, and to
$\sim 80$ per cent for the AGN case. Similarly to the DSC mass
fraction in the left panel, the BCG+DSC fraction is higher for relaxed
clusters than for un-relaxed ones.

The same cut at $r_{500}$ is applied to the simulated clusters after
computing the surface brightness maps.  Then, a SBL is applied
to the maps to identify the ICL, whose mass is defined as the stellar
mass within those pixels that are brighter than the surface brightness limit. As discussed
in Section \ref{map}, both luminosity and mass of star particles are
smoothed in the same way, so that each pixel is assigned mass and
luminosity consistently. 

In Fig. \ref{fig:ICLf_M}, we show the ICL mass fraction from the SBL
method. Upper left, upper right and lower left panels show the ICL
mass fraction obtained using three different surface brightness limits: $\mu_V > 25.5, 26.5,
27.5$, respectively. The symbols and lines have the same meaning as in
Fig.  \ref{fig:ICLf_L}. Obviously, using a higher SBL, fewer pixels are
selected as belonging to the ICL component, thus providing lower ICL
mass fractions. This trend with the SBL is stronger for the AGN
simulations; this reflects the fact that the galaxies in the AGN set are
less concentrated than in the CSF set. The ICL fractions for the AGN set
weakly decrease with halo mass, while the CSF clusters show no such
trend.  The decrease in the AGN set is in contrast with the increasing
trend found in several observations, as reported in the Introduction.
For the choice of $\mu_V > 26.5$, the ICL fraction is $\sim 30 -
20$ per cent for the AGN set, from low to high halo masses, and $\sim
15$ cent for the CSF set.
Those values are in line with those found in simulations by
\cite{Rudick2011} and by several authors from observational data
\cite[e.g.][]{Zibetti2005, Krick2007}.

In the lower right panel of Fig. \ref{fig:ICLf_M}, we show the BCG+ICL
mass fraction obtained with the SBL method using the reference value
for the surface brightness limit. The mass of the BCG is computed as described in Section
~\ref{map}.  The mass fractions in the BCG+ICL obtained with the SBL
method agree well with the BCG+DSC fractions obtained with the
dynamical method (see the right panel of Fig. \ref{fig:ICLf_L}) for
both sets of simulations.  This means the two methods only differ in
the way they separate the diffuse and the BCG components. With $\mu_V
= 26.5$, the SBL method assigns more star particles to the BCG than
the dynamical method does. 

With both simulations, we produce a higher BCG+DSC/ICL fraction than the
observation results of \cite{Gonzalez2007} (see left panel of
Fig. \ref{fig:ICLf_L} and lower left panel of Fig. \ref{fig:ICLf_M}).
This is a well known problem of numerical simulations, which is also
found in similar analyses by \cite{Planelles2013} and
\cite{Puchwein2010}.  A too large mass of BCGs in simulations may not
fully explain the discrepancy. In fact, as shown in
Figs \ref{fig:ICLf_L} and \ref{fig:ICLf_M}, including AGN
feedback in our simulations increases the mass fraction of BCG+ICL,
despite the reduction of the BCG mass. A possible explanation for the
too large stellar mass fraction in the BCG+DSC/ICL may lie in the limited
numerical resolution of our simulations, which should produce too
fragile galaxies that are easily disrupted within the cluster
environment. On the other hand, we also point out that the conversion
of galaxy luminosities to masses in observational data could also be
affected by systematic uncertainties. In their analysis,
\cite{Gonzalez2007} assumed a constant $M/L_I=3.6$ to convert observed
luminosities into stellar masses for the ICL and for the whole galaxy
population. Recent works \citep[e.g.][]{ Conroy2012, Conroy2013,
  Spiniello2013} show that the stellar IMF and, therefore, the
resulting $M/L$ value could depend on the velocity dispersion,
i.e. the mass of the galaxy. In particular, $M/L$ is shown to be a
growing function of the galaxy mass.  In this case, using a fixed
mass-to-light ratio to convert luminosities in masses could lead to an
under-estimate of the BCG mass with respect to the rest of the cluster
galaxy population, and, thus, to an under-estimate of the stellar mass
fraction in the BCG+ICL. This under-estimate will affect more
significantly the most massive clusters rather than less massive ones,
which are dominated by the BCG in their total cluster stellar mass. In
order to provide a rough estimate of the impact of such an effect, we
assume BCGs to be characterized by the top-heavy IMF by
\cite{Salpeter1955}, while the bulk of the cluster galaxy population
following a \cite{Chabrier2003} IMF. In the extreme case of most
massive clusters, this effect can increase the BCG+ICL stellar mass
fraction by a factor of up to $\sim 1.6$. Clearly, a more precise
estimate of this effect would require a more detailed analysis, which
is beyond the scope of this paper.

\subsection{Comparing the two simulation sets and the two methods}
\label{sec:md}

\begin{figure*}
\includegraphics[width=1.0\textwidth]{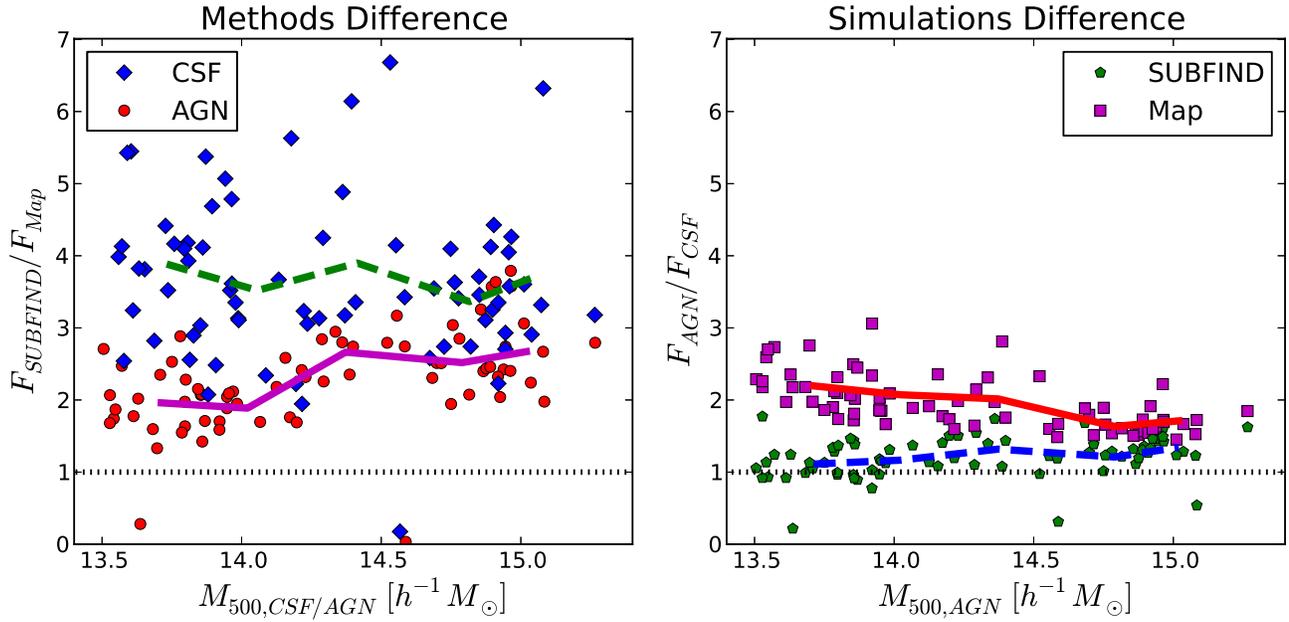}
\caption{Left panel: the ratio between the DSC and ICL fractions obtained
  with the two methods as a function of $M_{500}$. The blue solid
  squares denote the CSF simulations, while the red solid circles
  denote the AGN simulations. The dashed green and solid magenta lines
  show the mean value of blue squares and red circles,
  respectively. Right panel: ratio of the DSC/ICL fractions between the
  AGN and the CSF simulations, as a function of halo mass $M_{500,
    AGN}$ (we use the mass found in the AGN set). Magenta squares and
  red solid line refer to the ICL obtained with the SBL method; green
  pentagons and blue dashed line are for the DSC obtained with the
  dynamical method.}
\label{fig:DF_SM}
\end{figure*}

It is clear from Figs \ref{fig:ICLf_L} and \ref{fig:ICLf_M} that the
two simulation sets produce significantly different DSC/ICL mass
fractions for both methods. To make a detailed comparison between the
two simulation sets, we first match the two halo catalogues. Since each
simulated Lagrangian region contains few clusters, a halo in a CSF
simulation is matched to its counterpart in the AGN simulation when
the distance of their centres is less than 200 $\Kpc$ and their halo
mass difference is within $50$ per cent. In case there is no match
we simply exclude the halo from comparison, while if we find more than
one match we choose the nearest pair.

In the left panel of Fig. \ref{fig:DF_SM} we quantify, for each
simulation set, the difference in the DSC and ICL fraction obtained
with the two methods.  We show in the figure the ratio between the DSC
fraction within $r_{500}$ obtained with the dynamical method and the
ICL fraction obtained from the SBL method with $\mu_V > 26.5$.  The
difference between the results obtained using the two methods is
smaller for the AGN simulations, but even in this case the
dynamical-motivated DSC fraction is higher by a factor of 2 than the
SBL-based ICL, and the difference increases weakly with halo mass.
Within the CSF simulations the difference between the two methods
rises to a factor of $\sim 3-4$.  In \cite{Rudick2011} it is reported
that different methods can change the ICL fraction up to a factor of
2. This is consistent with our results with AGN feedback, while we
obtain an even higher difference with the CSF simulations.

In the right panel of Fig. \ref{fig:DF_SM} we show, for each method,
the ratio of the DSC or ICL fractions between the two simulation sets,
as a function of $M_{500}$ measured in the AGN set.  For both methods
AGN simulations give higher fractions than CSF simulations, by a
factor of $\sim 1.3$ for the dynamical method and by $\sim 2$ for the
SBL method. The fraction from the dynamical method is less affected by
the inclusion of AGN feedback than the fraction from SBL method.


\begin{figure*}
\includegraphics[width=1.0\textwidth]{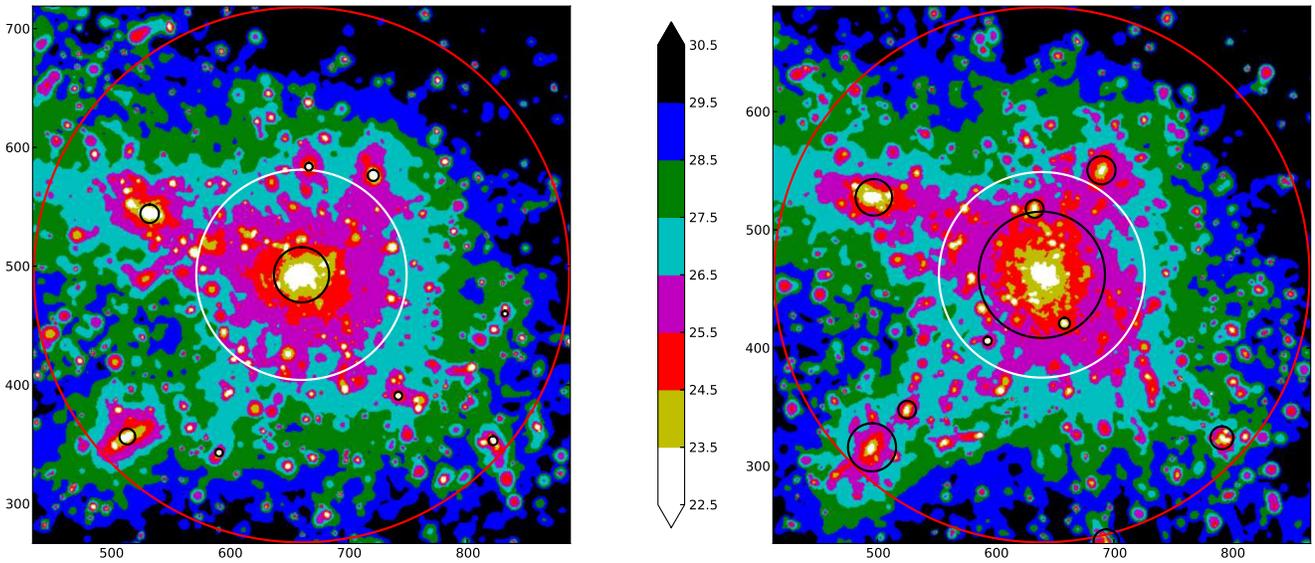}
\caption{Projected surface brightness maps for one cluster with
  $M_{500} \approx 8 \times 10^{14} / M_{\sun} h^{-1}$ in the two
  simulation sets. The left panel shows the CSF simulation, while the
  right panel shows the corresponding AGN simulation. Colour levels
  follow apparent magnitude per square arcsec, as specified in the colour
  side-bar. The standard $\mu_V=26.5$ surface brightness level
  corresponds to the level separating the cyan and magenta regions.
  The black circles show the radii of the BCG (the central one) and of
  satellite galaxies provided by {\small SUBFIND}. The red circle marks
  $r_{500}$. The smaller white circle shows the radius of the BCG as
  computed with {\small MAP} (Section~\ref{map}).}
\label{fig:MD_D1}
\end{figure*}

To understand the origin of these differences, we first check the
radii of the BCG as found with the {\small SUBFIND} and {\small MAP}
algorithms.  In Fig.
\ref{fig:MD_D1}, we show the surface brightness map (in rest-frame
$V$-band apparent magnitude per $\mathrm arcsec^2$) of one specific
cluster with $M_{500} \approx 8 \times 10^{14} h^{-1} M_{\sun}$.  The
left panel gives the map for the CSF simulation, while the right panel
is for the AGN one. Colour coding follows the curves of equal surface
brightness given by the colour bar.  We over-plot circles corresponding
to $r_{500}$ (red circle), to the radii of the BCG and of other
satellite galaxies from {\small SUBFIND} (black circles) and to the BCG radius
estimated with {\small MAP} (white circle).  As for {\small SUBFIND}, circles have a
radius equal to twice the half-mass radius of the stars that are
assigned to that galaxy. As for {\small MAP}, we use a surface brightness limit of $\mu_V = 26.5$
and the procedure introduced in Section \ref{map} (the largest radius
for which more than half the pixels overtake the SBL).  Clearly the
sizes of the two BCGs differs significantly, especially in the CSF
case (left panel of Fig. \ref{fig:MD_D1}). The difference is smaller
for the AGN case, but it still amounts to about a factor of 2. These
trends neatly reflect the varying discrepancy of the DSC and the ICL in the
CSF and AGN simulations, as seen in Fig. \ref{fig:DF_SM}. The small
black circles in Fig. \ref{fig:MD_D1} show the satellite galaxies'
size from {\small SUBFIND}, which is also twice the half-mass
radius. The sizes 
of satellite galaxies in the AGN case also agree better between the
two methods.

\subsection{The inferred surface brightness limit from the Dynamical Method}

\begin{figure*}
\includegraphics[width=1.0\textwidth]{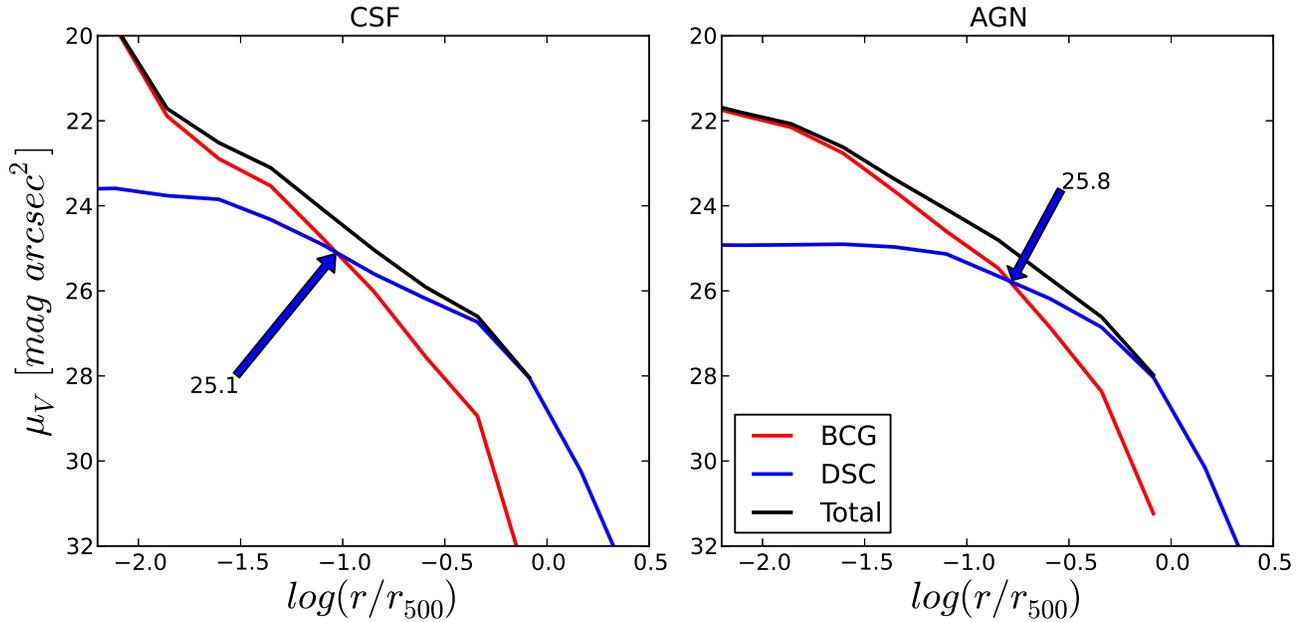}
\caption{Surface brightness profiles contributed by the stars assigned
  to the BCG (red lines) and to the DSC (blue lines) by the dynamical
  method from the same cluster as Fig. \ref{fig:MD_D1}. Black lines
  give the total surface brightness 
  profile. Arrows highlight the points where the BCG and DSC curves
  cross.  Left and right panels show the CSF and AGN simulations,
  respectively.}
\label{fig:apmp}
\end{figure*}

\begin{figure*}
\includegraphics[width=1.0\textwidth]{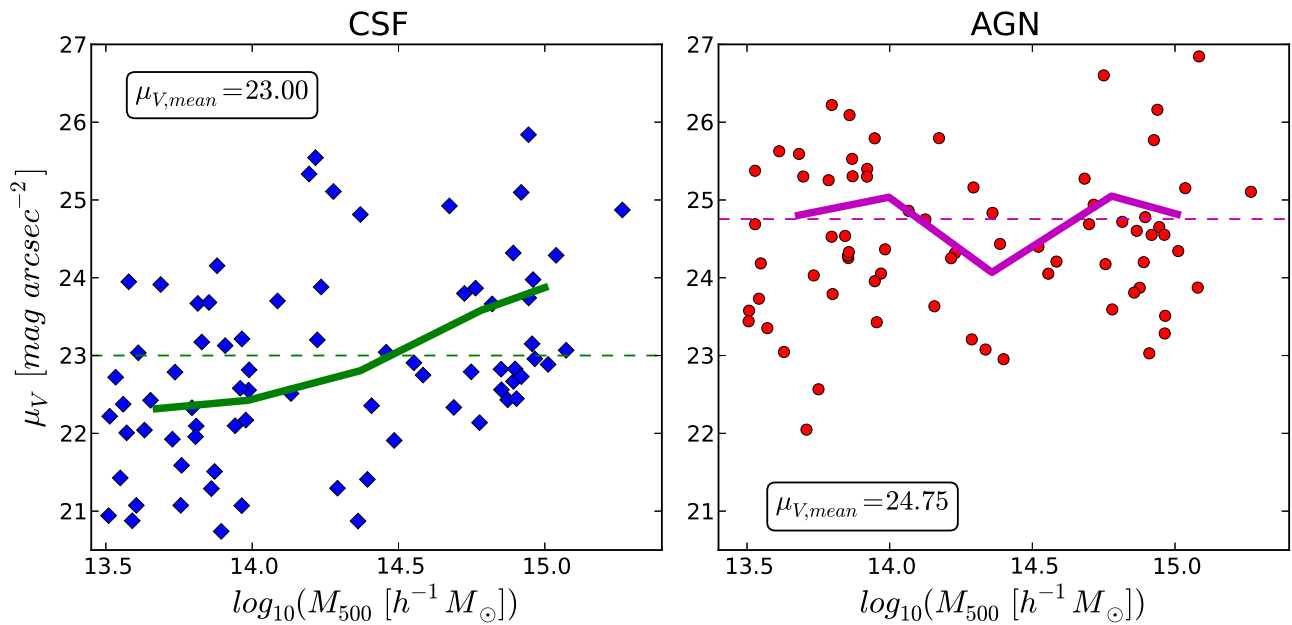}
\caption{Surface brightness values for all clusters at the point where the surface
  brightness curves of the BCG and DSC cross, as a function of $M_{500}$.
  The left panel is for the CSF simulations, and the right panel is for the
  AGN simulations. The solid thick lines in the two panels give
  average value in mass bins, while the horizontal thin dashed lines
  mark averages for all clusters with $M_{500} \geq 13.5$.}
\label{fig:appsp}
\end{figure*}

\begin{figure*}
\includegraphics[width=1.0\textwidth]{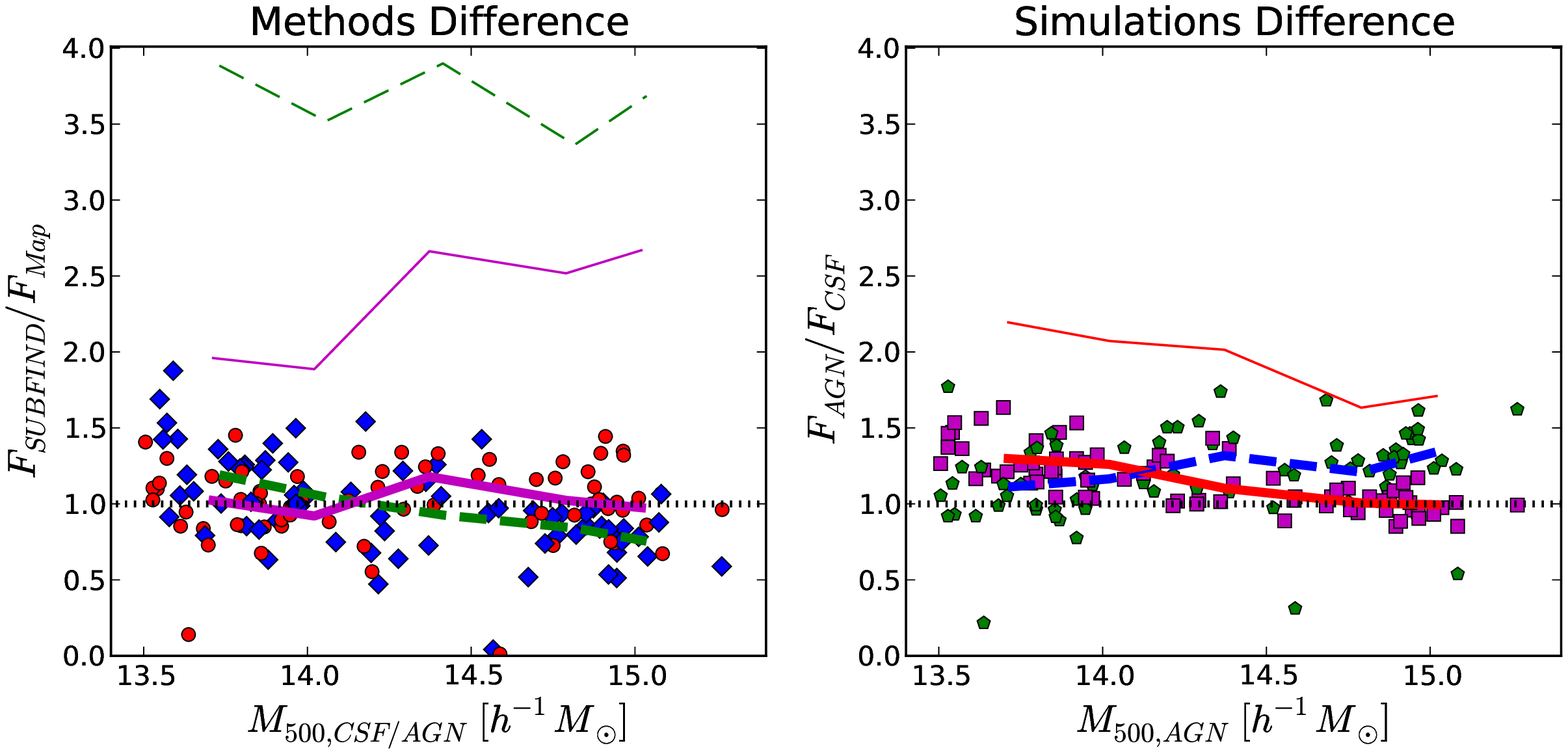}
\caption{As in Fig. \ref{fig:DF_SM}, we show, as a function of
  $M_{500}$, the ratio between the DSC and ICL fractions obtained with the
  two methods (left panel), and the ratio of the DSC/ICL fractions between
  the AGN and CSF set (right panel). In this case, the ICL fractions are
  obtained in {\small MAP} using SBLs of $\mu_V
  =23.00$ and $\mu_V=24.75$, as in Fig. \ref{fig:appsp}, for the CSF
  and AGN set respectively.  For an easier comparison with
  Fig. \ref{fig:DF_SM}, we show results obtained with the standard
  26.5 SBL as thin, green, 
  dashed (CSF) and thin, solid, magenta (AGN) lines in the left panel,
  and as thin, red, solid line (SBL method) in the right panel.}
\label{fig:FDSMsap}
\end{figure*}

The connection between galaxy size, both the DSC and ICL can be further
illustrated as follows. We plot in Fig.~\ref{fig:apmp} the radial
profile of surface brightness of the same cluster of
Fig. \ref{fig:MD_D1} for all stellar particles (black lines) and for
stars assigned by the 
dynamical method to the BCG (red lines) and to the DSC (blue
lines). AGN feedback, as implemented in these simulations, has a
strong effect on the BCG \citep[see][for a detailed
analysis]{Cin2013}, decreasing not only its final stellar mass but
also its compactness. Based on Fig.~\ref{fig:apmp}, we argue that the
surface brightness at which the profiles of the BCG and DSC cross
identifies a suitable value of the surface brightness limit to be used in the {\small MAP} algorithm to
best reproduce results of the dynamical method. Then, for each of our
simulated clusters we construct the ICL and BCG surface brightness
profiles and compute the value of surface brightness corresponding to
their crossing. These crossing values are shown in
Fig.~\ref{fig:appsp} for all the clusters in both simulation sets;
thick solid lines give mass-dependent averages, dashed lines show the
average value for all clusters with $\log (M_{500}/{\rm h^{-1}
  M}_\odot)> 13.5$.  Crossing values of $\mu_V$ display a large
scatter but are typically much brighter than $\mu_V=26.5$, the value
commonly used in observations to separate the BCG and ICL components. For
the CSF simulations, the $\mu_V$ crossing values are rather bright,
with an average of $\mu_V=23$ and a tendency to be fainter for more
massive clusters. For the AGN simulations their value is 
$\mu_V\simeq 24.75$ and show no trend with the cluster mass. This can
be understood as an effect of AGN feedback that makes BCGs more
extended and larger in size (Fig.~\ref{fig:MD_D1}), while lowering
their total stellar mass \citep[see also][]{Cin2013}.

We then use the above average crossing values, $\mu_{V}=23.00$ and
$\mu_{V}=24.75$ for the CSF and AGN cases, in the {\small MAP} algorithm to separate
out the ICL from the BCG. This leads to an increase in the ICL fraction that
becomes, as expected, similar to the corresponding DSC fraction.  In
the left-hand panel of Fig.~\ref{fig:FDSMsap} we show, as in
Fig.~\ref{fig:DF_SM}, the ratio between the DSC and ICL fractions for
the two sets of simulations. For comparison, average values from
Fig.~\ref{fig:DF_SM} are reported as dashed lines. Using the
crossing values of the SBL, the difference between two methods decreases
dramatically for both simulation sets, with average values raising
only for small clusters in the CSF set, for which a constant crossing
value of the surface brightness limit is clearly a poor fit (Fig.~\ref{fig:appsp}). Even the
scatter is modest, most fractions differ by less than a factor of 2
from the average values. The right panel shows the ratio between the ICL
fractions for two simulation sets, the thin red line giving the value
found for a SBL of $\mu_V=26.5$.  The difference between the two
simulation sets is now lowered by adopting the two new SBL values.
The difference in these ICL fractions between relaxed and un-relaxed
clusters is similar to earlier results.

This consistency check demonstrates that we understand the origin of the
different results provided by the two methods for the two simulation
sets. However, the dependence of the crossing value of
SBLs upon physics tells us that fainter surface brightness crossings
which have a better agreement with observations, are derived when
AGN feedback is included. This indicates that these simulations
with AGN feedback are headed in the right direction. The
large scatter in both simulations suggests that also in observations, a
single SBL applied to all BCGs could be too simplistic. We
do not go as far as suggesting that a SBL, say, of $\mu_V=24.75$ mag
arcsec$^{-2}$ should be used in observations to best reproduce the
results of a dynamically motivated algorithm to identify the DSC. 
However, in the future, it could be possible to apply this method to
infer the amount of the DSC once simulations are able to give
convergent and realistic properties (both masses and sizes) of BCGs.

\subsection{Velocity dispersions from the BCG and DSC}

\begin{figure*}
\includegraphics[width=1.0\textwidth]{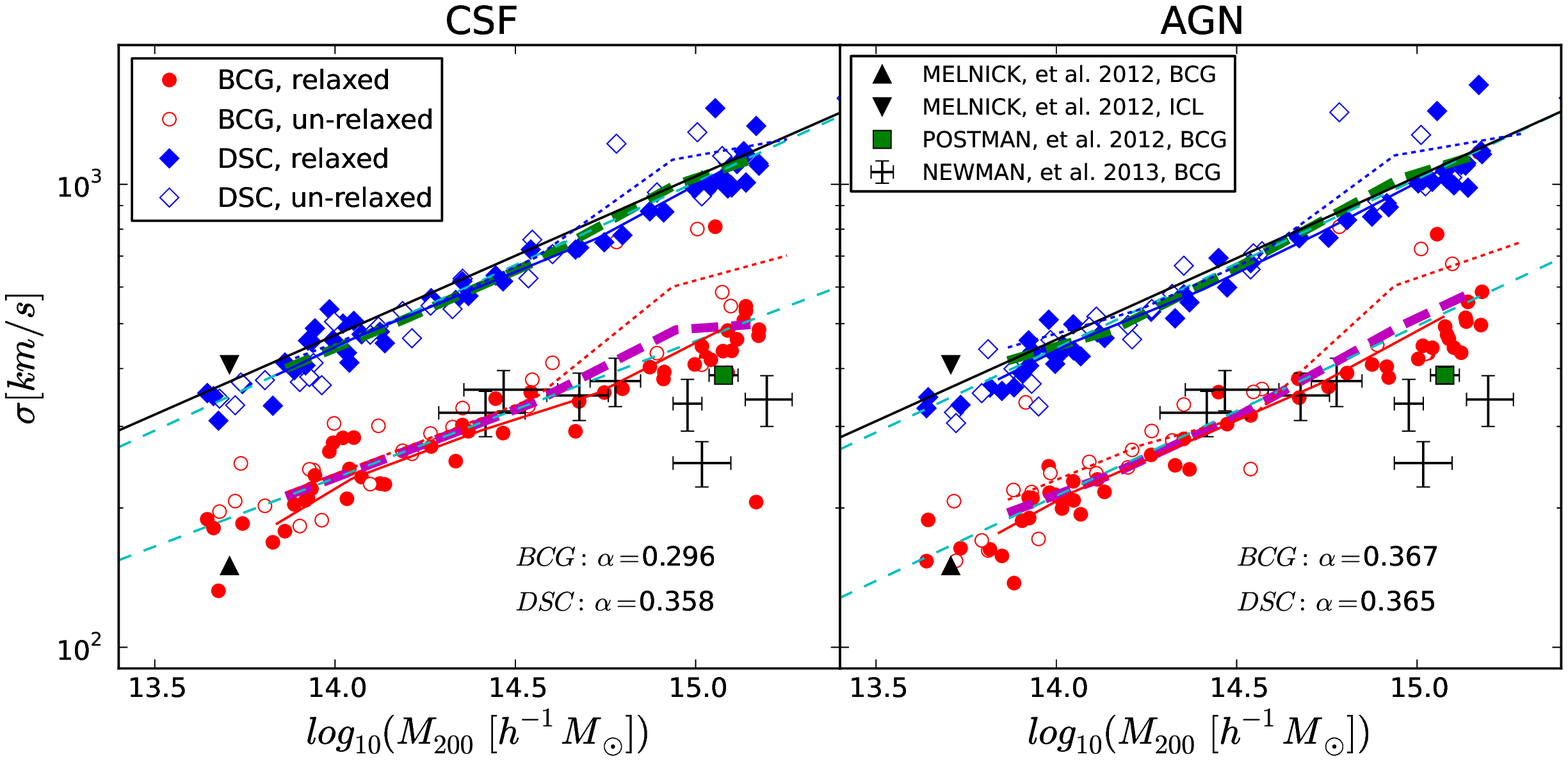}
\caption{Velocity dispersions of BCGs and DSC as a function of halo
  mass $M_{200}$. Red circles and blue squares indicate BCGs and DSC,
  respectively. Thick dashed magenta and green lines show the
  corresponding binned mean values. Filled and open symbols denote
  relaxed and un-relaxed clusters, with the mean value showed by solid
  and dotted lines of the same colour. The two dashed cyan lines are
  the linear fitting results, with their slopes $\alpha$ reported in
  the lower right side of each panel.  The black line shows the halo
  velocity dispersion fitting. Black triangles, filled green square
  and crosses with errorbars report observational results, according
  to the legend shown in the right panel.}
\label{fig:vd}
\end{figure*}

Since the DSC is the dynamical result of cluster evolution and is
naturally separated from the BCG using velocity information provided
by {\small SUBFIND}, it is interesting to investigate its velocity dispersion
$\sigma_{DSC}$, as well as $\sigma_{BCG}$.

From the observational side, there are very few cases in which
kinematic information on the ICL is available. \cite{Melnick2012}
investigated the dynamical information of the ICL in RXJ0054.0-2823,
an X-ray identified cluster at $z = 0.29$. They found
that its BCG has a velocity dispersion of $\sigma \approx 150 \vel$,
while the ICL velocity dispersion is $\sim 408 \vel$. They also
found a steep increase in the velocity dispersion with radius
qualitatively similar to what is observed by \cite{Coccato2011} in
Hydra-I. This supports that the double Maxwellian velocity profile
fitting method is a natural way of separating the two components
\citep[see also][]{Toledo2011}. 

In Fig. \ref{fig:vd}, we present the velocity dispersions of BCGs
and DSC from {\small SUBFIND} as a function of $M_{200}$. Those $\sigma$ for
both the BCG and DSC are consistent with the finding in \citet{Dolag2010}.
Assuming a power-law 
of the form $\sigma \propto M^{\alpha}$, both components show a
good fit with $\alpha \sim 0.36$ (except the BCG in the CSF set). This
value of $\alpha$ for the AGN simulations is very similar to that found by
\cite{Munari2013} from the analysis of these same simulation sets, in
which galaxies and subhaloes are used as tracers. Un-relaxed clusters
are under dynamic evolution, which results in a higher $\sigma$ for
both the BCG and DSC. That is why open symbols lie on the upper envelope
of the mean. By definition the DSC stars should be freely floating in
the gravitational potential of the cluster, and therefore should
essentially have a velocity dispersion similar to that of the cluster
as a whole (see the black fitting line). The velocity dispersion of the
DSC is higher than that of the BCG in the same halo, as expected. The
ratio is about a factor of 2 independent of halo mass. The values of
velocity dispersion of the BCG and ICL found by \cite{Melnick2012} are in
good agreement with our results (see the point-up and point-down black
triangles in Fig. \ref{fig:vd}), thus demonstrating that our
simulation captures the dynamical processes which lead to the formation
of the ICL and to the dynamical diversity of the diffuse component
with respect to the BCG component. The BCG velocity dispersion from
\cite{Newman2013} and \cite{Postman2012} also shows a good match to our
results. Note that we use a mean of the BCG velocity dispersion quoted
from table 6 of \cite{Newman2013} with errors from last radius
bin. 

\subsection{Physical properties of the BCG and ICL/DSC}

In this section, we analyse the difference between luminosity
and mass fractions associated with the ICL. Then, we focus on stellar
metallicities and ages of the BCG and of the ICL/DSC components
identified with the {\small SUBFIND} and {\small MAP} algorithms.

\begin{figure*}
\includegraphics[width=1.0\textwidth]{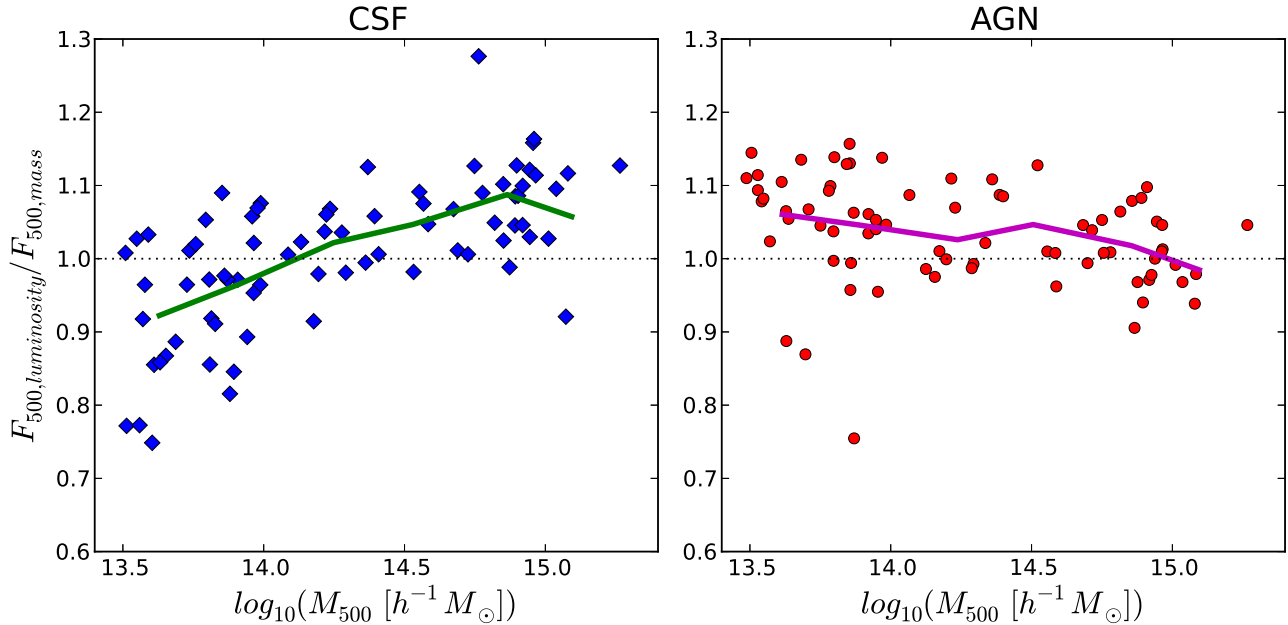}
\caption{Ratio of the luminosity-weighted and mass-weighted ICL
  fractions obtained from the {\small MAP} algorithm, using the same
  SBL $\mu_V = 
  26.5$.  The solid green and magenta lines are the average of the
  blue squares and red circles symbols in mass bins,
  respectively. Left and right panels show results for the CSF and AGN
  simulations, respectively.}
\label{fig:MD_L}
\end{figure*}

Observations provide estimates of the luminosity fraction of the ICL,
while the fractions studied before are all in mass.  Therefore a
proper comparison with observational results requires to quantify the
difference between luminosity and mass fraction in the ICL. This is
conveniently done using the SBL method. In Fig. \ref{fig:MD_L}, we
show the ratio of the two ICL fractions defined above for the two
simulations sets (left: CSF; right:AGN). For the CSF simulations, the
ratio of luminosity to mass fractions increases with halo mass, from
$\sim 0.9$ to $\sim 1.1$, while the same quantity decreases for the
AGN set from $\sim 1.1$ to $\sim$ 1.

For the {\small MAP} algorithm, we adopt the reference value for the SBL,
$\mu_V = 26.5$, to define the ICL, 
while the BCG is defined with the aperture method described in Section
\ref{map}.  All ages and metallicities are mass-weighted averages,
which is different from luminosity-weighted observation results.
In this section we show results for the AGN clusters only for the sake of
brevity. Results 
for the CSF simulations are qualitatively similar with the difference
that, due to the lack of quenching of late cooling flows, BCGs show
unrealistically high metallicities, young ages and blue colours.

\begin{figure*}
\includegraphics[width=1.0\textwidth]{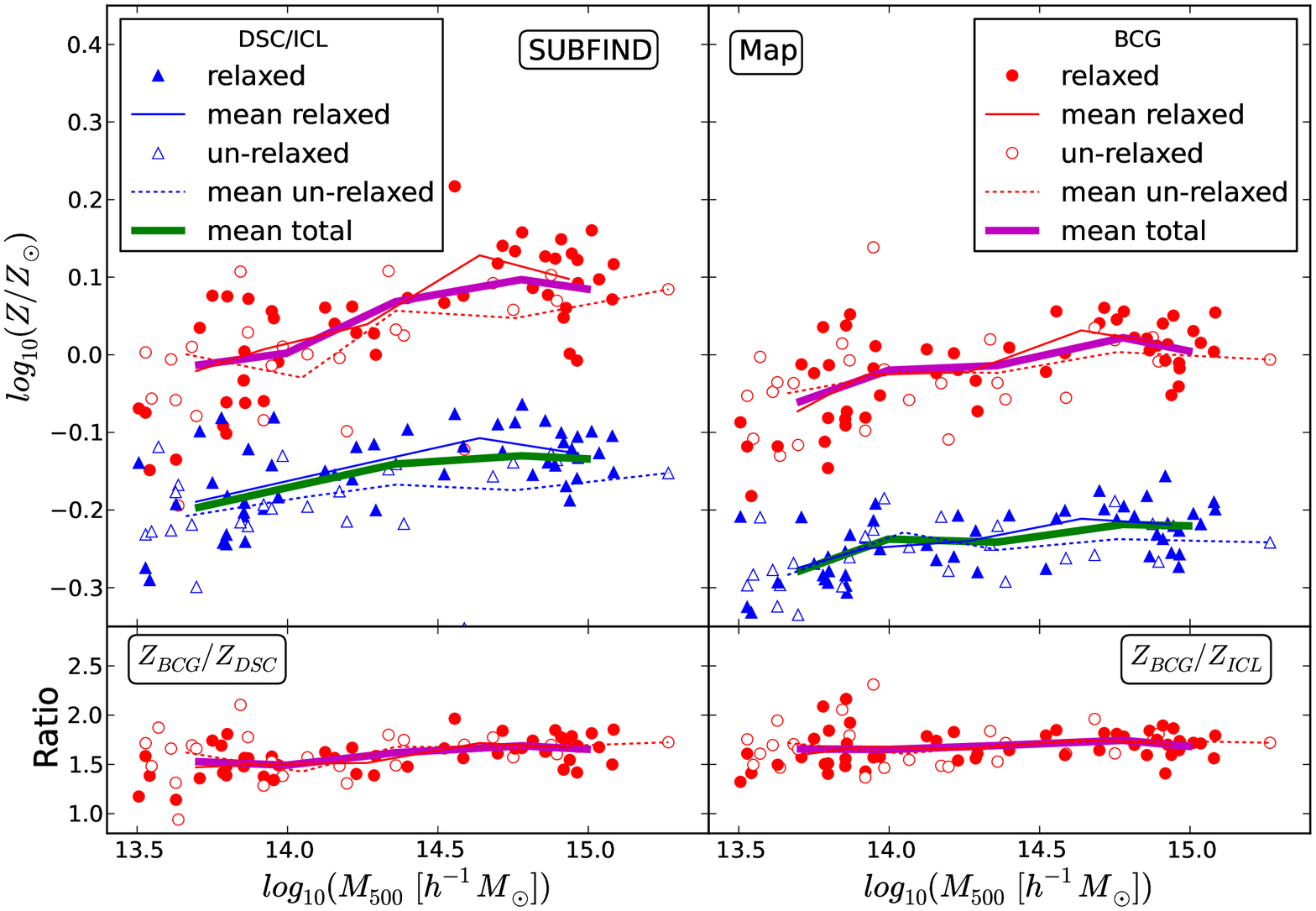}
\caption{Upper panels: average, mass-weighted metallicity of BCGs and
  DSC/ICL (left: {\small SUBFIND}; right: {\small MAP}).  Lower panels: ratios of
  metallicities of BCGs and DSC/ICL.  Filled and open symbols denote
  relaxed and unrelaxed clusters, respectively. Thick magenta and
  green lines give the corresponding averages in mass bins, thin red
  and blue continuous and dotted lines give averages for relaxed and
  un-relaxed clusters for BCGs and DSC/ICL, respectively.}
\label{fig:met}
\end{figure*}
\begin{figure*}
\includegraphics[width=1.0\textwidth]{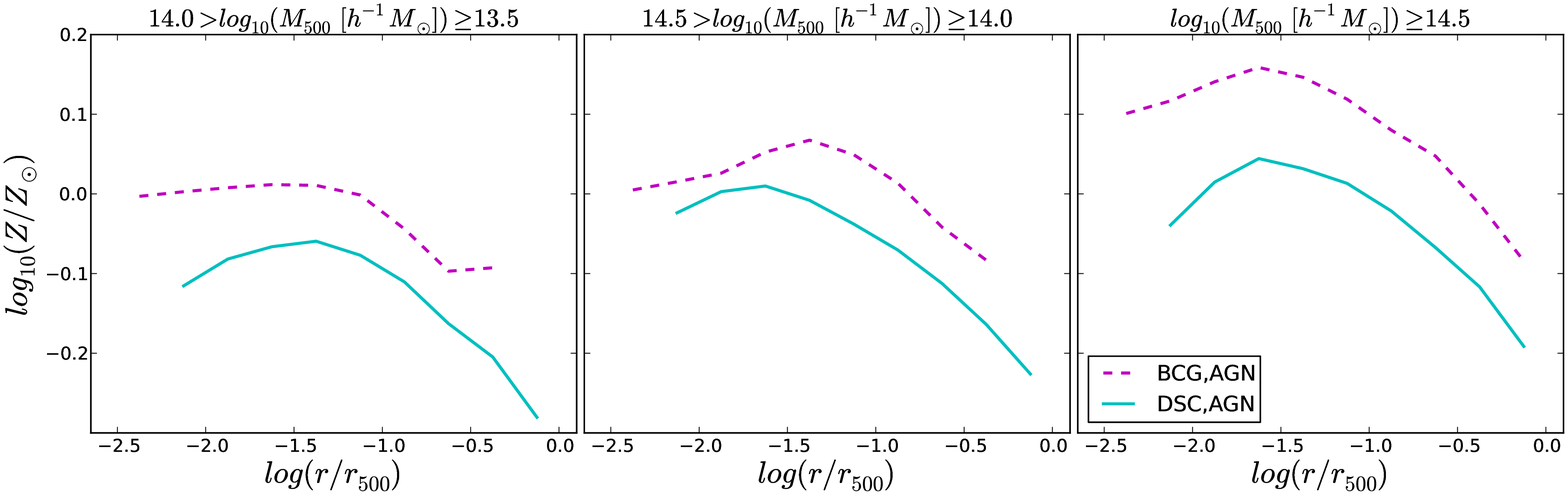}
\caption{Average metallicity profiles of the BCG (purple dashed line)
  and DSC (solid cyan line) stars from the dynamical method,
  as a function of radius (normalized to $r_{500}$), for clusters
  belonging to three mass-limited subsamples, as reported in the label
  above the panels.} 
\label{fig:met_prof}
\end{figure*}

\begin{figure*}
\includegraphics[width=1.0\textwidth]{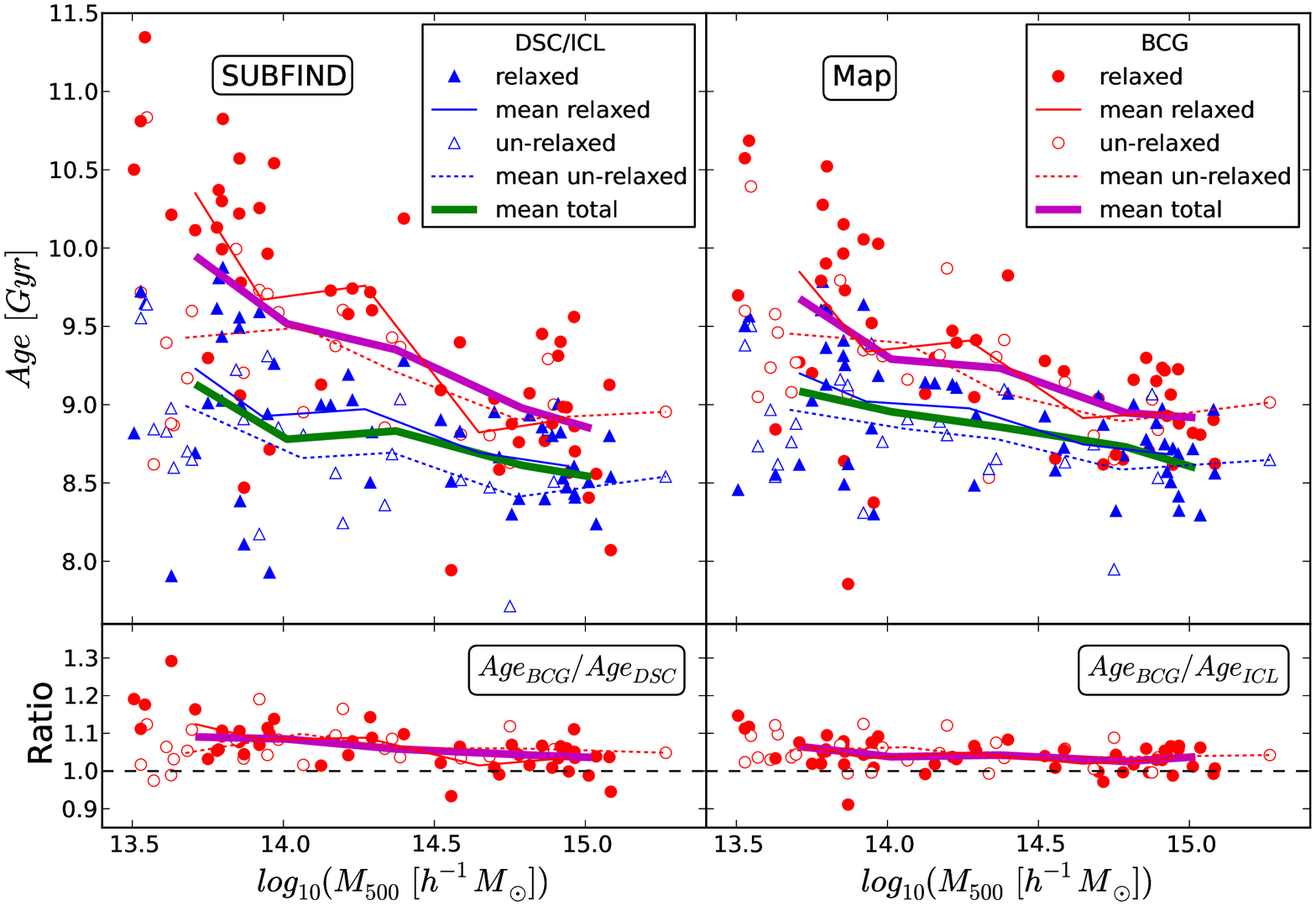}
\caption{Upper panels: average, mass-weighted age of BCGs and the DSC/ICL
  (left: {\small SUBFIND}; right: {\small MAP}).  Lower panels: ratios of ages of the BCG
  and the DSC/ICL. The symbols are the same as in Fig.~\ref{fig:met}.}
\label{fig:age}
\end{figure*}

In Figs \ref{fig:met} and \ref{fig:age}, we show
mass-weighted metallicities and ages of star particles assigned to the
BCG and to the DSC/ICL for both methods. In the upper panels we report
the BCG and DSC/ICL metallicities or ages from dynamical (left) and SBL
(right) methods. The BCG metallicities (Fig.~\ref{fig:met}) are roughly
solar and show a weak increasing trend with halo mass. The DSC/ICL
metallicities have a similar trend but with a systematic lower
value. The dynamical method gives higher metallicities by 0.1 dex than the
SBL method in both components. Ages (Fig.~\ref{fig:age}) decrease
with halo masses for both BCGs and DSC/ICL. BCGs systematically have
older age ($\sim 0.6 $ Gyr) than the DSC/ICL. Using mean age,
\cite{Dolag2010} also found that BCGs are older ($\sim 0.8$ Gyr) than the
DSC. However, since the simulations analysed by Dolag et al. did not
include AGN feedback, their ages for both components are about 4 Gyr
younger than ours. Relaxed clusters show slightly higher metallicity
and older age in both the BCG and the DSC/ICL components, for both methods.

In the lower panels of the same figures, we show the ratios of
metallicities and ages of the BCG and the DSC/ICL components.  In both cases,
the metallicity in the BCG component is about $60$ per cent higher than
the DSC/ICL metallicity, and the age is $\sim 6$ per cent higher. The
difference between relaxed and un-relaxed clusters is negligible in
both metallicity and age ratios.

\begin{figure*}
\includegraphics[width=1.0\textwidth]{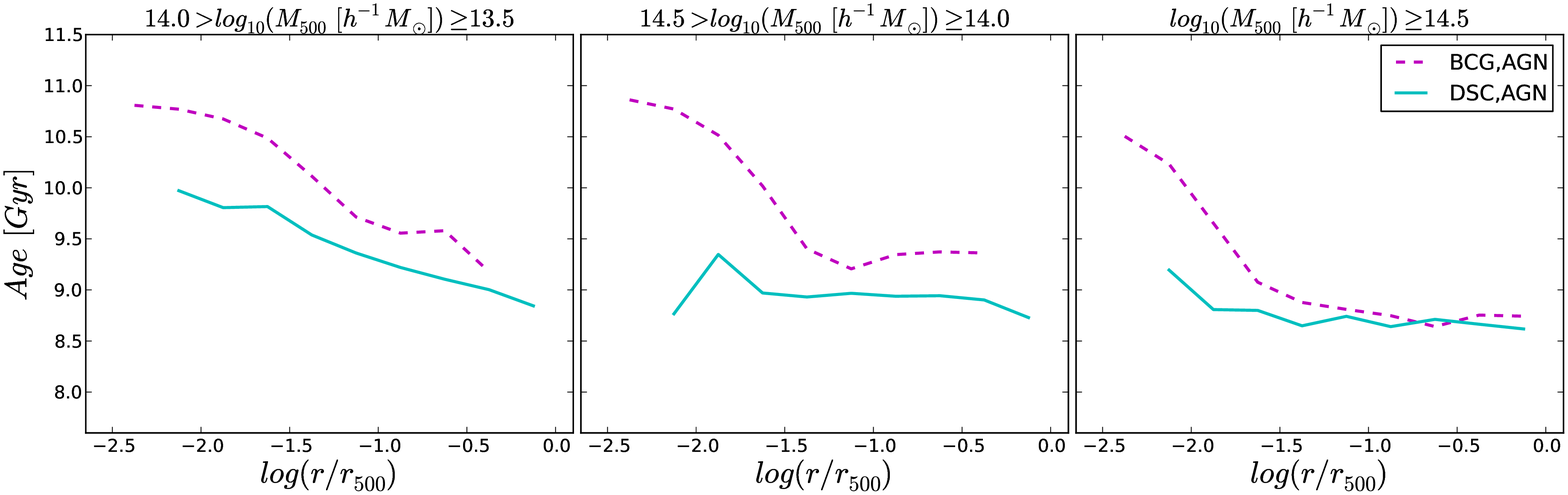}
\caption{Average age profiles of BCGs (purple dashed line) and the DSC
  (solid cyan line) stars as found by dynamical method, as a
  function of radius (normalized to $r_{500}$), in three different
  mass-limited subsamples as reported in the label above the panels.}
\label{fig:age_prof}
\end{figure*}

In Figs \ref{fig:met_prof} and \ref{fig:age_prof} we report
stacked metallicity and age profiles of the BCG and DSC from the
dynamical method obtained as follows. The cluster sample is divided
into three sub-samples according to their $M_{500}$ mass (see the
labels above the three panels).  For each sub-sample we distribute
star particles in 10 logarithmically equispaced bins of $r/r_{500}$,
ranging from $-2.5$ to $0$, and for each bin we perform a
mass-weighted average of metallicities or ages of all star particles
in that bin.

Metallicity profiles of the BCG stars (Fig.~\ref{fig:met_prof}) are higher
and flatter in the central regions, dominated by the BCG itself, while
there is a decreasing trend for the DSC stars. In the outer regions,
dominated by the DSC, the declining trend has a similar slope for both the
BCG and the DSC. Over all regions, the BCG metallicity profiles are higher
than the DSC profiles by $\sim 0.1$ dex. Metallicity profiles for both the
BCG stars and the DSC stars are clearly higher in the higher mass bins. On
the contrary, age profiles for BCGs (Fig.~\ref{fig:age_prof}) are
contributed by a much older population of stars towards the centre,
while the DSC age profiles are flatter, with only a slightly increasing
trend in age in the inner region. The low halo mass bin shows steeper
age profiles of those DSC stars compared to the other two mass bins.

\cite{Williams2007} found that intracluster population of stars in the
field of the Virgo cluster is dominated by low-metallicity stars 
with $[M/H] \la -1$, although 
this field appears to contain stars of the full range of
metallicities probed $-2.3 \leq [M/H] \geq 0.0$, and with
ages $\ga 10$ Gyr.
A similar result was obtained by \cite{Coccato2011} in the Hydra-I
(A1060) cluster using long slit spectroscopy Lick indices. However,
\cite{Melnick2012} claim that they found the ICL in their cluster to
be dominated by old metal-rich stars. Using semi-analytical models,
\cite{Contini2013} found the ICL metallicity $log (Z/Z_{\sun}) \sim
-0.2$, which agrees with our results. However, their BCG metallicity
is about 0.1 dex poorer than ours. \cite{Gabriella2012} also showed
that the stellar metallicities of the most massive galaxies would be
too low with respect to observational data. The BCG metallicity in our
simulations with AGN feedback is higher and closer to observations
compared to the predictions from the semi-analytical models by
\cite{Gabriella2012}. The mean DSC/ICL age from our AGN simulations is
$\sim 9$ Gyr, which is also comparable with observations. From the
metallicity and age profiles of the DSC stars, we predict that the DSC in
the inner cluster regions is mainly contributed by massive galaxies,
which are old and metal rich; meanwhile, further out, they are
slightly younger 
and more metal poor, which suggests stripping from dwarf galaxies, or
perhaps from the outer regions of spiral galaxies.


\section{Discussion and Conclusions}
\label{discussion}

In this paper, we analysed hydrodynamic simulations of galaxy clusters
and groups, with the aim of characterizing the diffuse component in
the distribution of stars, and the corresponding properties of the
intra-cluster light. This analysis has been carried out by applying
two methods for the identification of such diffuse intra-cluster
component. The first dynamical method is based on identifying stars
belonging to the main substructure with the {\small SUBFIND}
\citep{Springel2001} algorithm for the identification of
substructures. Besides separating the contribution of satellite
galaxies, our implementation of {\small SUBFIND} \citep{Dolag2010}
also includes 
a prescription for separating the BCG from the diffuse stellar
component by fitting the velocity dispersion profile with a
double Maxwellian distribution. The second, observationally-oriented
method is based on generating synthetic maps of $V$-band luminosity,
applying a photometric synthesis code \citep{Bruzual2003} to the
distribution of star particles. The intra-cluster light was then
separated from the light of the BCG by applying a surface brightness
limit (SBL) cut to the map (see Section \ref{map}).  These two methods
have been applied to two sets of simulations; a first one including
only the effects of cooling, star formation, chemical enrichment and
galactic outflows driven by SN-II feedback (CSF set); and a second one by
also including gas accretion on to supermassive black holes and the
ensuing effect of thermal AGN feedback (AGN set).

The main results of our analysis can be summarized as follows.

\begin{itemize}
\item The DSC mass fraction within $r_{500}$, obtained with the
  dynamical method, amounts to $\sim 40 - 50$ per cent for the CSF
  simulations, slightly increasing to $\sim 55$ per cent for the AGN
  case. Those value are higher than the corresponding ICL mass
  fraction obtained with the SBL method when a value of $\mu_V = 26.5
  \sbu$ is used for the surface brightness limit separating the two
  components: $\sim 15$ per cent for the CSF simulations and $\sim 20
  - 30$ per cent for the AGN simulations.  Moreover, the DSC/ICL
  fractions in the AGN set is higher than that of the CSF set by a
  factor of $\sim$1.5 and $\sim$2 when using the dynamical and the SBL
  methods, respectively. Conversely, the BCG+DSC/ICL mass fractions
  within $r_{500}$ are similar for the two methods, in both CSF and
  AGN simulations.

\item A comparison with available observational results shows too
  large a stellar mass fraction in the BCG+ICL found in our
  simulations. This discrepancy may be contributed both by the limited
  numerical resolution that makes simulated galaxies too fragile in
  the cluster environment, and by a possible underestimate in
  observations due to the assumption of a constant $M/L$ for the whole
  cluster galaxy population.

\item To understand the differences in the DSC/ICL fractions obtained
  with the two methods, we investigate the projection effects. We
  verified that the size of the region in the surface brightness map
  to be identified with the BCG is significantly larger than the size
  of the BCG identified by the dynamical method.  This explains the
  larger amount of mass in the ICL from the SBL method, with respect
  to that in the DSC from the dynamical method.

\item For each cluster, we computed the value of the surface
  brightness threshold that gives the same ICL mass fraction as the
  DSC one provided by {\small SUBFIND}. These values are $\mu_V\simeq 23$ and
  $24.75$ for the CSF and AGN simulations, respectively, thus much
  brighter than the values commonly adopted in observational studies,
  although with a large object-by-object scatter and a weak trend with
  mass in the CSF case.

\item The velocity dispersions $\sigma$ of both BCGs and the DSC show
  a tight scaling with the total cluster mass, $\sigma \propto
  M_{200}^{\alpha}$, with $\alpha \simeq 0.36$. At fixed $M_{200}$ the
  velocity dispersion of the DSC is larger than that of the BCG by
  about a factor of 2, consistent with the fact that the DSC is
  associated with the broader Maxwellian velocity distribution. These
  results are consistent with observational evidence for a larger
  value of the velocity dispersion of the DSC
  with respect to that of the BCG \citep[e.g.][]{Melnick2012}.

\item  The DSC and ICL components show
  similar ages and metallicities, but both lower than the BCG ones.
  Age and metallicity profiles of stars identified as belonging to the
  BCG or the DSC by the dynamical method significantly differ in the
  central parts, dominated by the BCG. The BCG stars lying in the central
  regions are older and more metal rich than the DSC stars.

\end{itemize}

The analysis of our simulations shows in general that the DSC in
galaxy clusters represents a distinct population 
with respect to the component associated with the BCG. These two
components not only have different kinematics, thereby hinting to a
different dynamical origin, but are also characterized by different
ages and metallicity. These results are generally in line with
observational evidence, thus demonstrating that simulations are able
to capture the basic phenomena leading to the generation of the
intra-cluster stellar populations. Still, a quantitative comparison
with observational data demonstrates the existence of a tension with
predictions from simulations. First of all, even with the inclusion of
AGN feedback, simulations still produce BCGs that tend to be more
massive than the observed ones \citep[see also][ and references
therein]{Cin2013}. As a consequence, identifying ICL in simulations
based on the same surface brightness limit adopted in observational
data can not lead to an identical comparison between real and
simulated clusters. On the other hand, observational measurements of
the stellar mass of the BCG are also prone to systematic uncertainties,
especially if they are based on photometric data, combined with
a suitable value for the stellar mass-to-light ratio, usually
assumed to be constant.

There is little doubt that fully exploiting the potential of the
diffuse intra-cluster stars as tracers of the past assembly and
star-formation histories in galaxy clusters requires a leap forward in
the observational characterization of this elusive
component. Furthermore, it also requires simulations to include
reliable descriptions of the key astrophysical processes which
regulate the evolution of this galaxy population.

\section*{Acknowledgements}
The authors would like to thank Mario Nonino for useful
discussions. All the figures in this paper are plotted using the
python matplotlib package \citep{Hunter:2007}.  Simulations have been
carried out at the CINECA supercomputing Centre in Bologna, with CPU
time assigned through ISCRA proposals and through an agreement with
the University of Trieste.  WC and MK acknowledge a fellowship from
the European Commission's Framework Programme 7, through the Marie
Curie Initial Training Network CosmoComp (PITN-GA-2009-238356).  PM
and SB acknowledge support from FRA2009 grant from the University of
Trieste. GDL acknowledges financial support from the European Research
Council under the European Community's Seventh Framework Programme
(FP7/2007-2013)/ERC grant agreement n. 202781.  This work is partially
supported by the PRIN/MIUR-2009 grant ``Tracing the growth of cosmic
structures'', by the PRIN-INAF 2010 grant ``Towards an italian network
for computational cosmology'' and by the PD51-INFN grant. KD
acknowledges the support by the DFG Cluster of Excellence "Origin and
Structure of the Universe".

\bibliographystyle{mnras}
\bibliography{bibliography}

\label{lastpage}
\end{document}